\documentclass{aa}

\usepackage[varg]{txfonts}
\usepackage{xcolor}
\usepackage[normalem]{ulem}
\usepackage{comment}
\usepackage[utf8]{inputenc}

\usepackage[symbol]{footmisc}

\usepackage{hyperref}
\hypersetup{
    colorlinks,
    linkcolor={blue!50!black},
    citecolor={blue!50!black},
    urlcolor={blue!50!black}
    }

\usepackage[encapsulated]{CJK}
\usepackage{ucs}
\newcommand{\orcid}[1]{\protect\href{https://orcid.org/#1}{\protect\includegraphics[width=8pt]{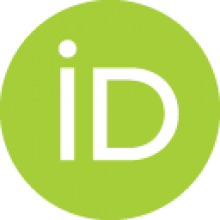}}}
\usepackage[utf8]{inputenc} 

\newcommand{\orcit}[1]{\protect\href{https://orcid.org/#1}{\protect\includegraphics[width=8pt]{orcid-ID.png}}}

\usepackage{comment}

\begin{document}

\makeatletter
\renewcommand{\linenumbers}{\relax}

\title{Revisiting 3C\,279 jet morphology with space VLBI at 26 microarcsecond resolution}

\author{
Teresa~Toscano \orcid{0000-0003-3658-7862}\inst{\ref{IAA}}$^{,\dagger}$
\and José L. Gómez
\orcid{0000-0003-4190-7613}\inst{\ref{IAA}}
\and Guang-Yao~Zhao
\orcid{0000-0002-4417-1659}\inst{\ref{Mpifr}, \ref{IAA}}
\and Rocco~Lico
\orcid{0000-0001-7361-2460}\inst{\ref{INAF},\ref{IAA}}
\and Antonio~Fuentes
\orcid{0000-0002-8773-4933}\inst{\ref{IAA}}
\and Tuomas~Savolainen
\orcid{0000-0002-8773-4933}\inst{\ref{aalto},\ref{aalto2},\ref{Mpifr}}
\and Jan~Röder
\orcid{0000-0002-8773-4933}\inst{\ref{IAA},\ref{Mpifr}}
\and Maciek~Wielgus
\orcid{0000-0002-8635-4242}\inst{\ref{IAA}}
\and Alexander B. Pushkarev\inst{\ref{crimea},\ref{moscow}}
\orcid{0000-0002-9702-2307}
\and Efthalia Traianou
\orcid{0000-0002-1209-6500}\inst{\ref{IWR}, \ref{Mpifr}}
\and Ai-Ling~Zeng
\orcid{0009-0000-9427-4608}\inst{\ref{IAA}}
\and Leonid~I.~Gurvits
\orcid{0000-0002-0694-2459}\inst{\ref{jive}, \ref{tud}, \ref{shao}}
\and Yuri Y. Kovalev
\orcid{0000-0001-9303-3263} \inst{\ref{Mpifr}}
\and Felix M. Pötzl
\orcid{0000-0002-6579-8311} \inst{\ref{greece}, \ref{greece2}}
\and Mikhail Lisakov
\orcid{0000-0001-6088-3819}\inst{\ref{PUCV}}
}

\institute{
Instituto de Astrof\'{i}sica de Andaluc\'{i}a-CSIC, Glorieta de la Astronom\'{i}a s/n, E-18008 Granada, Spain\label{IAA} $^{\dagger}$\email{ttoscano@iaa.es}
\and Max-Planck-Institut für RadioAstronomie, Auf dem Hügel 69, D-53121 Bonn, Germany\label{Mpifr}
\and INAF Istituto di RadioAstronomia, Via Gobetti 101, 40129 Bologna, Italy\label{INAF}
\and Aalto University Department of Electronics and Nanoengineering, PL 15500, 00076 Aalto, Finland\label{aalto}
\and Aalto University Metsähovi Radio Observatory, Metsähovintie 114, 02540 Kylmälä, Finland\label{aalto2}
\and Crimean Astrophysical Observatory, 298409 Nauchny, Crimea
\label{crimea}
\and Astro Space Centre of Lebedev Physical Institute, Profsoyuznaya 84/32, Moscow 117997, Russia \label{moscow}
\and Interdisziplin\"ares Zentrum f\"ur Wissenschaftliches Rechnen (IWR), Universit\"at Heidelberg, Im Neuenheimer Feld 205, 69120 Heidelberg, Germany \label{IWR}
\and Joint Institute for VLBI ERIC (JIVE), Oude Hoogeveensedijk 4, 7991 PD Dwingeloo, The Netherlands \label{jive}
\and Faculty of Aerospace Engineering, Delft University of Technology, Kluyverweg 1, 2629 HS Delft, The Netherlands \label{tud}
\and Shanghai Astronomical Observatory, Chinese Academy of Sciences, 80 Nandan Rd., Shanghai 200030, China \label{shao}
\and Institute of Astrophysics, Foundation for Research and Technology – Hellas, N. Plastira 100, Voutes GR-70013 Heraklion, Greece \label{greece}
\and University of Crete, Department of Physics \& Institute of Theoretical \& Computational Physics, 70013 Heraklion, Greece \label{greece2}
\and Instituto de F\'{i}sica, Pontificia Universidad Cat\'{o}lica de Valpara\'{i}so, Casilla 4059, Valpara\'{i}so, Chile \label{PUCV}
}

\titlerunning{Looking into 3C\,279 filamentary evolution}
\authorrunning{T.~Toscano et al.}

\date{Received \today / Accepted \today}

\abstract{We present observations of the blazar 3C\,279 at 22 GHz using the space VLBI mission \textit{RadioAstron} on 2018 January 15. Images in both total intensity and fractional polarization are reconstructed using RML method implemented in the \texttt{eht-imaging} library. The electric vector position angles are found to be mostly aligned with the general jet direction, suggesting a predominantly toroidal magnetic field, in agreement with the presence of a helical magnetic field. Ground-space fringes were detected up to a projected baseline length of $\sim 8$~G$\lambda$, achieving the angular resolution of around 26~$\mu$as. The fine-scale structure of the relativistic jet is found in our study extending to a projected distance of $\sim 180$~parsec from the radio core.
However, the filamentary structure reported by previous \textit{RadioAstron} observations of 2014 is
not detected in our current study. We discuss potential causes for this phenomenon, together with a comparison using public 43~GHz data from the BEAM-ME program, showing a significant drop in the jet's total intensity. The optically thick core is observed with a brightness temperature of $ 1.6 \times 10^{12}$~K, consistent with equipartition between the energy densities of the relativistic particles and the magnetic field. 
This yields an estimated magnetic field strength of 0.2 G.}


\maketitle

\section{Introduction}
\label{sec:intro}


The blazar 3C\,279 (1253$-$055) at redshift $z=0.536$ \citep{Marziani_1996} is a bright, highly polarized and well-monitored source, one of the first objects to provide evidence of rapid structure variability \citep{knight1971} and apparent superluminal motions in compact active galactic nuclei (AGN) jets \citep{whitney1971, cohen1971}. Since then, the structure of the jet in 3C\,279, comprised of a compact core and a jet extended from sub-parsec to kiloparsec scales, has been thoroughly studied across the whole accessible electromagnetic spectrum, showing high variability and frequent flares. The core of 3C\,279 has a high brightness temperature at centimeter wavelengths \citep[e.g., $\gtrsim 10^{12}$~K in][]{Kovalev_2005}, as well as a high fractional linear polarization in the jet \citep[$\gtrsim 10\%$, ][]{Pushkarev_2023}. In the extended jet, different components show a wide range of apparent speeds (from a few to $\sim$ 20c) and a bulk Lorentz factor in the range of $\Gamma \sim 10$--40 \citep{bloom2013, Jorstad_2017, Weaver_2022}, suggesting the presence of features associated with propagating shocks or instabilities. 


Studying the formation, acceleration or collimation of relativistic jets, together with their connection with the supermassive black hole (SMBH), continues to be one of the most extensive pursuits in modern astrophysics \citep[see, e.g.,][and references therein for recent reviews]{Boccardi_2017, Blandford_2019}. Different theoretical models have been proposed to interpret this connection, mainly represented by two scenarios (not mutually exclusive), which suggest that either relativistic AGN jets are driven by the conversion of the rotational energy of the black hole (BH) to Poynting flux via magnetic field lines, which are attached to the BH ergosphere \citep{Blandford_1977}, or that these jets are instead driven by strong magnetic fields anchored onto the BHs accretion disk \citep{Blandford_1982}. 

Very Long Baseline Interferometry (VLBI) is a powerful technique that provides great angular resolution, enabling us to resolve and analyze the fine structure of extragalactic jets on (sub-)parsec scales. By observing at higher frequencies or extending the baseline length of the array elements, VLBI enables the study of jet features located at $\sim$ $10^{3}$ Schwarzschild radii from the central engine for a number of AGN sources. 
In particular, polarimetric VLBI observations play a crucial role in studying extragalactic jets, providing essential insights into jet physics and the configuration of magnetic fields \citep[e.g.,][]{gomez_2001, Gomez_2016, Gabuzda-2021, Gomez_2022}.

Different monitoring campaigns, like BEAM-ME \citep{Jorstad_2017} and MOJAVE \citep{Homan_2021,Lister_2021} have successfully and extensively observed these jets at millimeter and centimeter wavelengths, respectively. Particularly for 3C\,279, the MOJAVE program, monitoring at 15 GHz, has reported in this source maximum apparent speeds up to $\sim 20c$, consistent with a jet oriented at a small angle to the line of sight and moving with a Lorentz factor around $\Gamma \sim 13$  \citep{Lister_2019}. Similarly, BEAM-ME at 43 GHz has observed component speeds ranging from $\sim 4c$ to $\sim 21c$ \citep{Jorstad_2017}. Other findings using the Very Long Baseline Array (VLBA) have reported a dramatic change in the trajectory of a superluminal component (C4), as well as apparent projected speed and direction, moving with a Lorentz factor $\sim$ 15 suggesting that the trajectory change is a collimation event occurring at around 1 kpc (deprojected) from the core \citep{Homan_2003}.

The Event Horizon Telescope (EHT) composed of mm-wavelength radio telescopes distributed over the globe provided the first images of BH shadows with the sharp resolution of $\sim 20 \,\mu$as \citep{EHTC2019a, EHTC2022a}. 

The EHT has also imaged 3C\,279, presenting the first 1.3 mm VLBI images at the extreme angular resolution of 20 $\mu$as \citep{kim_2020}. Their results reveal a multicomponent inner jet morphology with the northernmost component elongated nearly perpendicularly to the direction of the jet, separated from the other component by $\sim$ 100 $\mu$as. 
This is interpreted as either a broad resolved jet base or a spatially bent jet, with low apparent (T$_{\text{b, obs}}$ $\lesssim$ 10$^{11}$ K) and intrinsic brightness temperatures (T$_{\text{b, int}}$ $\leq$ 10$^{10}$ K).


The \textit{RadioAstron} space VLBI mission \citep{Kardashev_2013}, specifically designed to study of ultra compact regions in space objects 
\citep{2020AdSpR..65..705K}, operated between 2011 and 2019. Led by the  Astro Space Center of Lebedev Physical Institute and the Lavochkin Association, this mission featured a 10~m radio telescope on board the \textit{Spektr-R} satellite, equipped with dual polarization receivers operating at 0.32 GHz (P band), 1.6 GHz (L-band), 4.8 GHz (C-band), and 22 GHz (K-band). 
With an apogee close to 350,000 km, the observations with \textit{RadioAstron} enable the imaging of blazar jets in both total and linearly polarized intensity with unprecedented resolution, that is, of the order of few tens of microarcseconds \citep[e.g.,][]{2015A&A...583A.100L, Gomez_2016, Giovannini_2018, 2020A&A...641A..40V, Bruni_2021, Gomez_2022, Savolainen_2023}.

The launching, collimation, and magnetic field properties of AGN jets have been studied as part of AGN Imaging Key Science Programmes (KSP), complemented by the AGN survey on the brightness temperature of their cores \citep{lobanov_2015a, Kovalev_2016}. 
The \textit{RadioAstron} Polarization KSP has constantly collected data throughout the whole duration of the space VLBI mission, successfully probing the regions of the jet closest to the core, as well as their magnetic field, for a sample of the most energetic blazars. Results of the \textit{RadioAstron} mission can be found for several sources such as BL Lac \citep{Gomez_2016}, 3C\,84 \citep{Savolainen_2023} or OJ\,287 \citep[][Traianou et al. 2025, submitted]{Gomez_2022,Cho_2024}, among others. The blazar 3C\,279 was among the sources included in the \textit{RadioAstron} Polarization KSP,
yielding the first robust detection and imaging of a filamentary structure of the jet \citep{Fuentes_2023}.

Results from \citet{Fuentes_2023} show, for the first time, the presence of filaments in the 3C\,279 radio jet. These bright filaments are suggested to be compressed regions with enhanced gas and magnetic pressure, likely as a result of Kelvin-Helmholtz (KH) instabilities. Moreover, these filaments exhibit periodic structures and brightness variations attributed to Doppler boosting. This periodicity also matches features observed at 7\,mm \citep{Jorstad_2017, Weaver_2022} and likely originates from an elliptical surface mode in a kinetically dominated cold jet. 
In addition, the source polarization signature discussed in their work indicates the presence of a toroidal magnetic field rotating clockwise along the flow with a pitch angle of $-45^{\circ}$ and an estimated bulk Lorentz factor of $\sim 13$ \citep{Jorstad_2005, Fuentes_2023}. 

\begin{table}[ht]
\centering
\caption{List of observing radio telescopes.}
\begin{tabular}{clcc}
\hline
Array & Station & Code & Diameter (m) \\
\hline
       & Mopra               & MP     & 22 \\
LBA    & Narrabri (ATCA)     & AT     & 22 (×6) \\
       & Ceduna              & CD     & 30 \\
       &                     &        &    \\

       & Tamna               & KT     & 21 \\
KVN    & Ulsan               & KU     & 21 \\
       & Yonsei              & KY     & 21 \\
       &                     &        &    \\
       & Tianma              & T6     & 65 \\
       & Badary              & BD     & 32 \\
       & Hartebeesthoek      & HH     & 26 \\
       & Metsähovi           & MH     & 14 \\
EVN    & Medicina            & MC     & 32 \\
       & Effelsberg          & EF     & 100 \\
       & Svetloe             & SV     & 32 \\
       & Zelenchukskaya      & ZC     & 32 \\
       &                     &        &    \\
SRT    & Spektr-R            & RA     & 10 \\
\hline
\end{tabular}
\label{tab:telescopes}
\end{table}
Here, we present an investigation of 3C\,279 at 22~GHz using \textit{RadioAstron} observations of this source performed on January 15, 2018, to explore the jet structure, polarization and time evolution of the filamentary structure previously detected in the 2014 observations \citep{Fuentes_2023}. 

In the following sections, we present the total intensity and polarization images of 3C\,279 at 22 GHz obtained with the space-VLBI mission \textit{RadioAstron}, obtained with the data described in Sec. \ref{sec:Obs}. Moreover, a supplementary study is introduced in different subsections showing the relation with gamma-rays (Sec.~\ref{subsec:gamma}), tests on synthetic data (Sec.~\ref{subsec:synth}), brightness evolution from the public BEAM-ME 43~GHz VLBA data (Sec.~\ref{subsec:BEAM}), as well as a comparison with the previous \textit{RadioAstron} data from 2014 (Sec.~\ref{subsec:comparison}) and a brightness temperature analysis (Sec.~\ref{subsec:Bright_temp}).

\section{Observations and data analysis}
\label{sec:Obs}

The observations (project code: GG083B) were carried out on 2018 January 15 using 14 ground-based radio telescopes and the \textit{RadioAstron} space radio telescope (see Table \ref{tab:telescopes})
at a frequency of 22.22~GHz (1.3~cm, K-band). 
The polarization calibration was carried out with the source 3C\, 279 (see Sec.~\ref{subsec:imaging}). 
The observations were carried out in two sessions with two distinct antenna arrays, spanning from 00:49:50 UTC on January~15 to 00:30:45 UTC on January~16 2018, with a four-hour gap (from 10:00 to 14:00 UTC) during which no observations were made.

\begin{figure}
    \centering
    \includegraphics[width=\columnwidth]{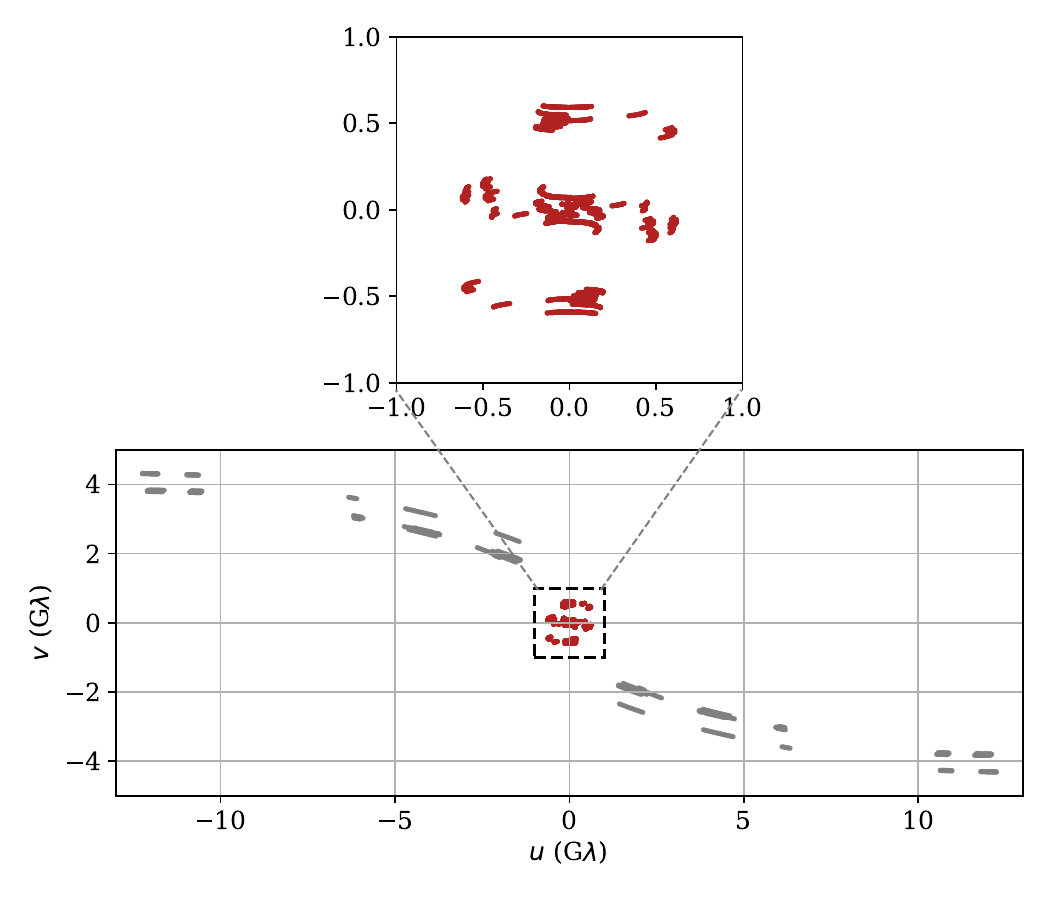}
    \caption{Coverage in the uv-plane
    of the fringe-fitted interferometric visibilities of 3C\,279, observed by \textit{RadioAstron} at 22~GHz on 2018 January 15. Top panel shows only the ground baselines, while bottom panel shows all baselines, with RadioAstron ones in grey.}
    \label{fig:uv-cover}
\end{figure}

The data were recorded in both left (LCP) and right (RCP) circular polarization, using 4 intermediate frequencies (IFs) with a recorded bandwidth of 16~MHz each (i.e., a total bandwidth of 64 MHz)
for all stations, except BD, SV, ZC, and \textit{RadioAstron}, which operated with 2 IF bands (i.e., a total bandwidth of 32 MHz). On the ground, data were sampled using 2-bit sampling, while RadioAstron used 1-bit. This difference affects the overall sensitivity and data rates of the observations.The $uv$-coverage of the observations is displayed in Fig.~\ref{fig:uv-cover}.

\subsection{Data reduction}

The raw data were correlated at the Max-Planck-Institute for Radio Astronomy using the \textit{RadioAstron} dedicated version of the DiFX software correlator \citep{Bruni_2016}. Initial data calibration and processing were performed using the NRAO Astronomical Image Processing System (AIPS; \citealt{Greisen_2003}), employing the \texttt{ParselTongue} interface \citep{2006ASPC..351..497K}, calibrating first the ground array and then adding the space antenna.



The gap during the observation naturally led to splitting the data into two subsets for separate calibration. For each subset, the same calibration process was applied, correcting for the parallactic angle and opacity for the ground stations. 

An a priori calibration of the correlated visibility amplitudes was performed based on the system temperatures and gain curves recorded at each station. Since six ground antennas (AT, BD, CD, SV, T6, ZC) did not provide system temperature information, we used the default (average) nominal values provided by each station and modulated them by the antenna’s elevation at each scan. We then solved for residual single- and multi-band delays, phase offsets, and phase rates by fringe-fitting the data using AIPS's task FRING. 

In the initial stage, we calibrated first only the ground array, and a global fringe fit with an exhaustive baseline search was performed on the ground array with a solution interval of 60 seconds. EF and T6 served as the reference antennas for the first and second data subsets, respectively.

Once the ground array was calibrated, we accounted for  
the lower sensitivity of the longest projected baselines by adopting different solution intervals (60s and 120s) and combining IFs and polarizations. Afterwards, to enhance fringe detection sensitivity on the longer \textit{RadioAstron} space-ground baselines, we coherently phased the ground array, aligning the phases of the individual antennas to improve the signal-to-noise ratio and sensitivity of potential fringe detections  \citep{Gomez_2016, Bruni_2016, Fuentes_2023}.


With a signal-to-noise ratio cut-off of 4, reliable ground-space fringes were detected in the first subset at projected baseline distances of up to around 8 Earth diameters 
providing a maximum angular resolution of 26 $\mu$as with a beam of 188 $\times$ 26 $\mu$as and orientation $\sim$ 21 degrees from North to East, transverse to the jet direction. 
For the second subset, with the longest baselines, we detect fringes only at the AT-RA baseline with projected length up to almost 13 Earth diameters, meaning no closure phases could be used. Alas, with a signal-to-noise ratio below 4, these detections were not used for the imaging. 

Finally, we solved for the antennas’ bandpasses and corrected the delay differences between polarizations using the AIPS task RLDLY. Both subsets were at last unified into a single one using AIPS's task DBCON.

After the calibration was completed, the data were carefully edited using Difmap \citep{Shepherd_1994, Shepherd_2011} in order to flag outliers that could potentially introduce noise and/or create artifacts during imaging.

The instrumental polarization (D-terms) were iteratively obtained using the \texttt{eht-imaging} software library \citep{Chael_2016, Chael_2018}, in the same way as previous work by \citet{Fuentes_2023}, by using 3C\,279 as a calibrator. Absolute calibration of the electric position vector angle (EVPA) was applied using single dish observations from the Effelsberg telescope, on the same day as our observations, of value 25.1 $\pm$ 1.8 degrees. The error in the EVPA calibration is estimated to be around 10 degrees.

For our particular dataset, before the imaging took place, we cut the edge channels, ending up with a bandwidth of 14.5 MHz for each IF. We also performed an initial phase-only self-calibration to a point source model for every point and coherently averaged the data in 120 second intervals using Difmap, a standard procedure to stabilize visibility phases in time while fully preserving the closure phase structure. We also checked that during the process of self-calibration the SNR of the data was sufficiently high in order to avoid generating spurious signal \citep{2021AJ....161...88P,Savolainen_2023}.

\subsection{Imaging strategy}\label{subsec:imaging}

\begin{figure*}[t]
    \centering
    \includegraphics[trim=0 0 2.1cm 0, clip,width=0.99\textwidth]{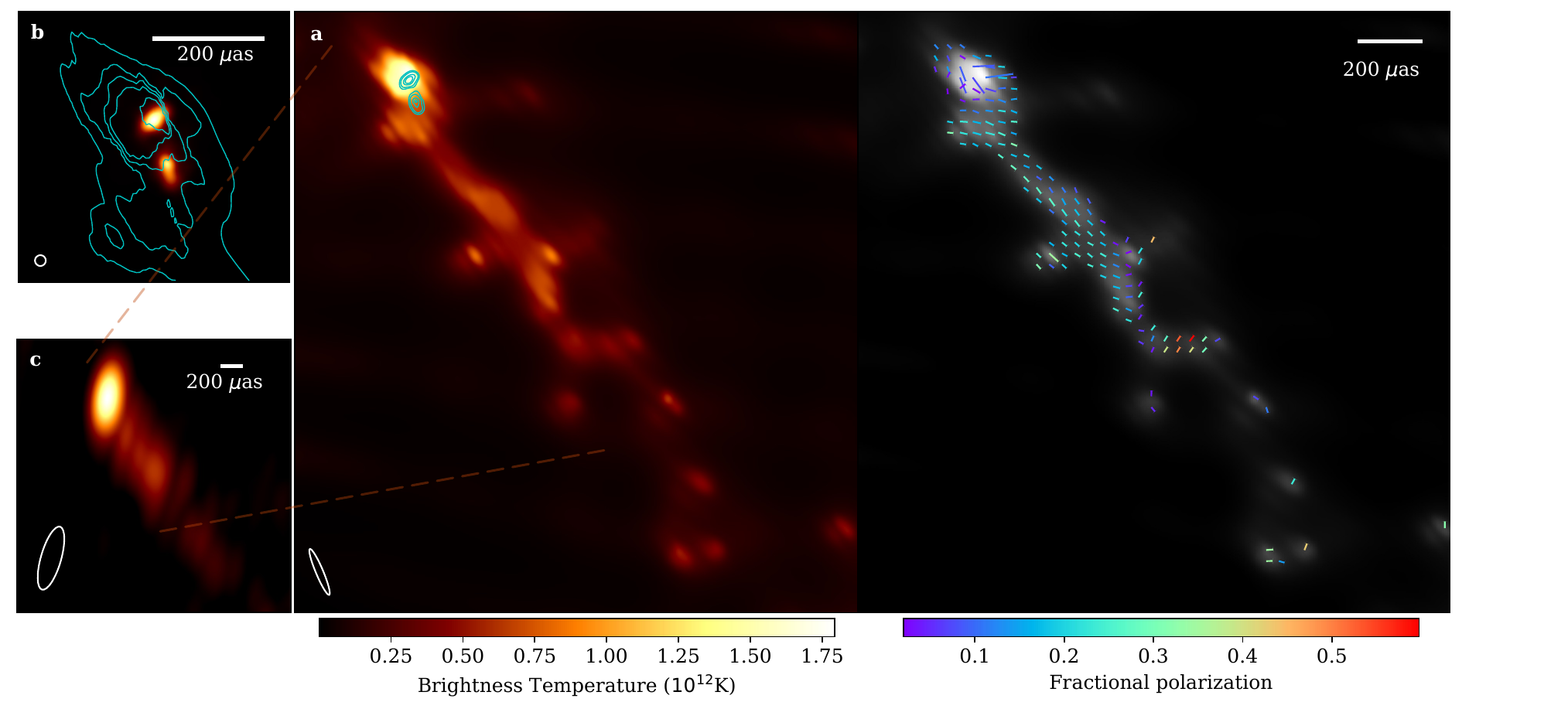}
    \caption{Jet structure of 3C\,279 revealed by \textit{RadioAstron}. a: Total intensity (left) and linearly polarized (right) \textit{RadioAstron} image at 22 GHz obtained on 2018 January 15. Both images show brightness temperature in color scale (\textit{afmhot} and grey scale, respectively), though the image on the right also shows the recovered electric vector position angles overlaid as ticks. Their length and color are proportional to the level of linearly polarized intensity and fractional polarization, respectively. EHT aligned image from panel b is also displayed in contours over total intensity image. b: The EHT image at 230 GHz obtained in April 2017 at 1:1 scale and aligned to our \textit{RadioAstron} image -in contours- using their respective pixel with maximum brightness. Contours are equally spaced in 7 levels within the data range. c The 43 GHz close-in-time image from BEAM-ME program from Feurbary 2018. White ellipses at the bottom-left corners show the convolving beams. Bottom colour bars refer only to information displayed on a.}
    \label{fig:RA_image}
\end{figure*}

Imaging was carried out using the \texttt{eht-imaging} software library, a regularized maximum likelihood (RML) method \citep{Chael_2016, Chael_2018}. While CLEAN algorithm \citep{Hogbom_1974,1995BAAS...27..903S} is widely used for VLBI image reconstruction \citep{Gomez_2022, Cho_2024}, 
RML methods can provide increased resolution over CLEAN, naturally incorporating both physical and prior constraints and showing better capabilities to super-resolve the source structure. Their superior resolution and fidelity make them ideal for sparse VLBI arrays. For instance, \texttt{eht-imaging} has been widely tested within the EHT collaboration at mm-wavelengths, and it also provides successful results for centimetre wavelengths and space VLBI experiments \citep{EHTC2019d,kim_2020,EHTC2022c,Fuentes_2023, Cho_2024}.

In general terms, RML methods try to obtain an image (I) in total intensity by minimizing an objective function J defined as follows:

\begin{equation}
    \hspace{1.2cm} J(I) = \sum_{\text{data terms}} \alpha_{D} \chi_{D}^{2} (I,V) - \sum_{\text{reg. terms}} \beta_{R} S_{R} (I), 
    \label{eq1}
\end{equation}

\noindent where \textit{D} and \textit{R} are a set of selected data products and regularization terms, respectively, and $\alpha_{D}$ and $\beta_{R}$ are hyperparameters that weight the contribution of the image fitting to the data terms $\chi^{2}_{D}$, and the image-domain regularization terms $S_{R}$, to the minimization of the previous equation. 

This imaging approach can use closure data products -closure phases (cphase) and log closure amplitudes (logcamp)- together with complex visibilities, providing an observable that is not affected by station-based errors 
in the image, while Difmap is only able to use closure quantities for self-calibration but not for imaging.
The significant number of radio telescopes participating in the observation allowed the use of closure quantities which proved to mitigate atmospheric phase corruption and gain uncertainties. 

The use of regularization terms also implies that certain features can be enhanced and imposed, such as flux density contraint (flux), smoothness between adjacent pixels (tv), compactness (l1) or pixel-to-pixel similarity to a prior image (mem, or 'simple' within the algorithm) \citep{Chael_2018}. Typical weighting values for both data terms and regularizers are either powers of 10, or factors of 100 as they are easier to interpret. Unlike fully Bayesian methods, RML techniques do not estimate the posterior distribution of the underlying image, but instead compute a single image that minimizes \ref{eq1}. 


Imaging the data requires several steps. First, in order to obtain a ground-only total intensity (Stokes I) image,  we flagged all baselines to \textit{RadioAstron} and imaged the data collected only by ground radio telescopes. Additional systematic uncertainty for each visibility,  1.5\% of the visibility amplitude, was added in quadrature to the thermal noise to account for non-closing errors. The pipeline was initialised with a Difmap CLEAN model of the data (see Appendix \ref{appendixa}), which provided better results than the typically used Gaussian prior. This CLEAN model was created using only the first subset of data, which was averaged over 120 seconds and underwent several rounds of cleaning and self-calibration at each data point. This subset contained primarily short and medium-length baselines, allowing for an initial estimation of the source structure. This way, it provides a better starting point for the \texttt{eht-imaging} imaging pipeline and, therefore, it helps the algorithm converge, preventing it from getting caught in local minima.

As mentioned above, we had poor a priori amplitude calibration caused by missing system temperature measurements for some ground antennas. This led us to perform an initial round of imaging using only closure quantities (closure phases and log closure amplitudes) to constrain the image likelihood via the mean squared standardized residual ($\chi^2$). In this step we used the following weights for data terms and regularizers: cphase: 100, logcamp: 100, flux:1e4, l1:10, mem:0, tv:1. This approach yielded values close to one (1.16 for complex visibilities, 1.00 for closure phases, and 0.78 for closure amplitudes), indicating a good fit to the data.

The total flux density in the final image was fixed to 10.2 Jy, a value measured as the average of the KVN baselines, the shortest ones available in our observations.

\begin{table*}[ht]
\centering
\caption{Summary of the main differences between 2014 and 2018 observations of 3C\,279 at 22 GHz with \textit{RadioAstron}.}
\label{tab:diff}
\begin{tabular}{c c c c c c c c }
\hline\hline
                                    Epoch & Flux (Jy) & Ground stations & Resolution ($\mu$as) & Filament detection & \multicolumn{1}{l}{\begin{tabular}[c]{@{}l@{}}Fractional \\ polarization\end{tabular}} & \multicolumn{1}{c}{\begin{tabular}[c]{@{}c@{}}On-source \\ time \end{tabular}} \\ \hline
\multicolumn{1}{l}{March 2014} & 27.2  $\pm$ 0.2               & 23                       &        
  27                   & yes                                               & $\sim$ 10$\pm$ 1\%         & 11\rm h 44 \rm min                    \\ 
\multicolumn{1}{l}{January 2018} & 10.2    $\pm$ 0.5           & 14                       &        26                 & no                                                 & $\sim$ 5 $\pm$ 1 \%        & 10\rm h 46 \rm min                      \\ \hline
\end{tabular}
\end{table*}
Each imaging iteration took the image reconstructed in the previous step blurred to the nominal resolution of the ground array (that is, 223 $\mu$as). 
Then, we did self-calibration fitting only phases to the closure-only image, and repeated the imaging but incorporating full complex visibilities. Here, the weighting of the data terms and regularizers are: cphase: 100, logcamp: 100, vis:20, flux:1e4, l1:10, tv:10, mem:1.


We then repeated the process, this time using the ground-array image as both the prior and the initial model for the algorithm, after blurring it to match the nominal resolution of the full array restored data (including \textit{RadioAstron} baselines.
That is, first imaging using only closure quantities (using cphase: 100, logcamp: 100, flux:1e4, l1:10, mem:0, tv:10), then applying self-calibration using only phases, and finally adding complex visibilities in the data term in several iterations, progressively increasing their weight in each of them (using cphase: 100, logcamp: 100, vis:50, flux:1e4, mem:10, tv:10 in the final iteration). We only included amplitude self-calibration in the final iteration step.
To ensure the robustness of the reconstruction, we also performed a parameter survey using different combinations of the regularizers (see Appendix \ref{appendixb}), finding a similar structure in all reconstructions.

Once the total intensity image was produced, we simultaneously estimated the D-terms iteratively for each station using the \texttt{eht-imaging} library, similar to other published works \citep{EHTC2021a, EHTC2024g, Fuentes_2023}.

\section{Results and discussion}
\label{sec:Results}

\begin{figure}
\centering
    \includegraphics[trim=0 3cm 0 3.5cm, clip, width=\columnwidth]{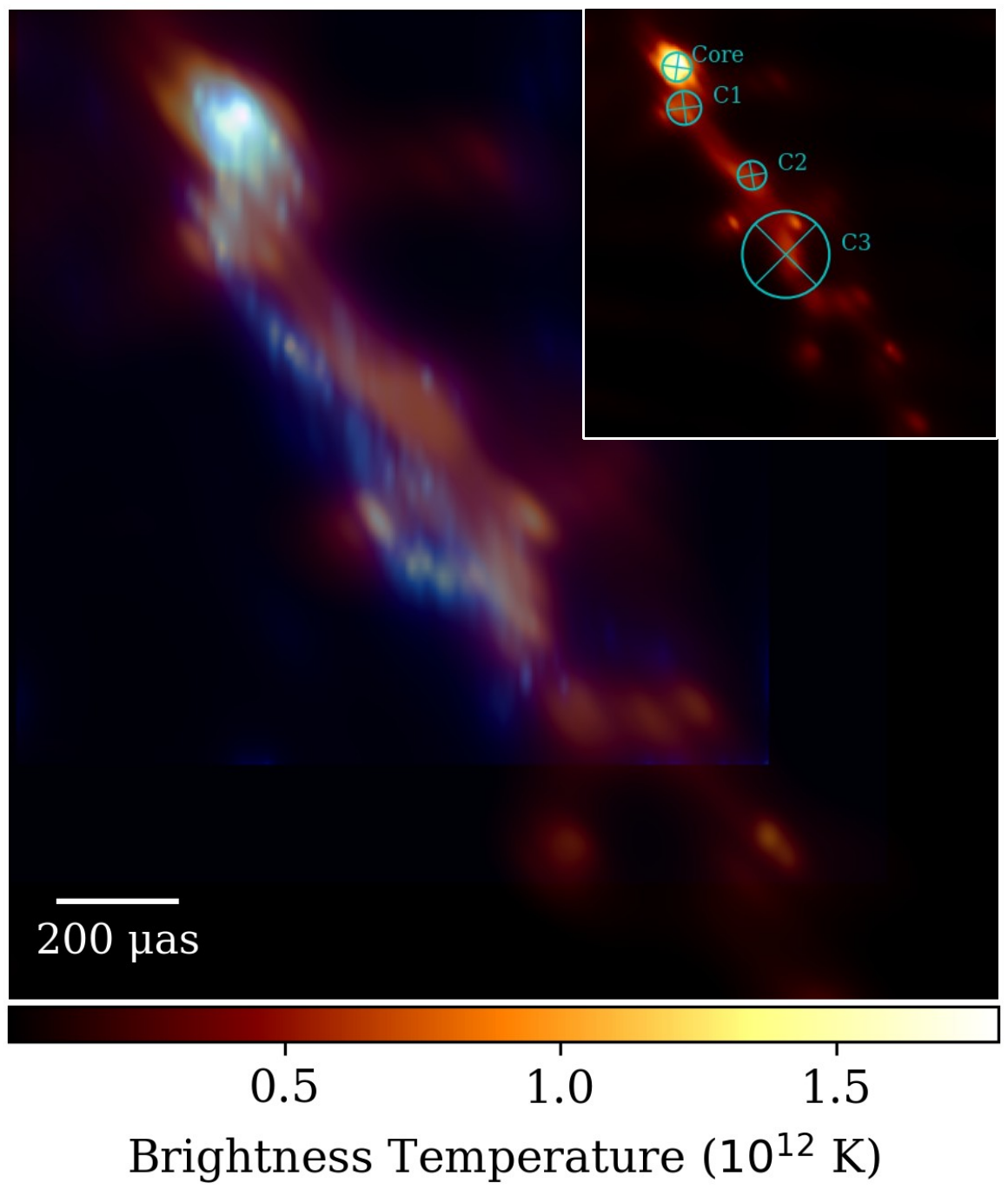}

    \caption[Comparison between 2014 and 2018 \textit{RadioAstron} images.]{Main image: a direct comparison between the 2014 and 2018
    images of 3C\,279 at 22~GHz with \textit{\textit{RadioAstron}}. The 2018 epoch is plotted in warm color scale \textit{afmhot} and aligned with the 2014 epoch \citep{Fuentes_2023}, in blue, using the brightest pixel. Top right image shows overlapped the Gaussian components from the model fitting process.
    The fifth component (E) is not shown since it is a large component necessary to take into account the diffuse emission.}
    \label{fig:comparison}
\end{figure}



In Fig. \ref{fig:RA_image}, the total intensity and linearly polarized images of 3C\,279 from 2018 January 15 are displayed, showing a field of view of about 2 x 2 $ \rm mas$, a beam of $26 \times 188 \, \mu \rm as$ and a total flux density of 10.2 Jy (see Sec. \ref{subsec:imaging}). The radio morphology of this source is known to contain a bright compact core and a thin jet \citep{Paterperley_1983}, like we also observe. The core is located at the northeast of the image, as the brightest region, followed by the jet's extended emission in an approximately narrow, straight line that ends with some diffuse emission. This 
morphology is consistent with previous observations and public data \citep{Lister_2018, kim_2020, Fuentes_2023}.

The core region displays a bend towards the South of almost 90 degrees connecting with the rest of the straight emission, similar to the structure found in 2014. 
At around 1 mas from the core, the extended jet shows a gentle curve towards the southwest, coincident with the region where the filament was observed in 2014 but slightly further away from the core (see also Fig. \ref{fig:comparison}). No filamentary structure has been detected for our 2018 epoch, though. 
However, the similarities in structure hint towards the existence of a possibly more complex structure that is not showing in this particular epoch, maybe due to the sparsity of the uv coverage and the decrease in brightness of the source (see Sec. \ref{subsec:synth} for more detail).

Fig.~\ref{fig:RA_image} also allows us to study the polarized structure of the source. The EVPA direction (plotted as overlayed ticks that vary in length and color, which are proportional to the level of linearly polarized intensity and fractional polarization, respectively) appears to be mostly following the direction of the jet, suggesting that the magnetic field in the jet has a predominant toroidal component in the plane of the sky (relativistic effects are assumed not to strongly influence the perpendicular direction between electric and magnetic fields direction). 

At the core, EVPAs show a less organized distribution. 
As for the rest of the emission, it mostly shows an EVPA orientation parallel to the jet direction, as in 2014 observations \citep{Fuentes_2023}, therefore in agreement with a dominant toroidal magnetic field.


Fig.~\ref{fig:RA_image} also shows BEAM-ME (43~GHz) close-in-time and EHT (230~GHz) 2017 observations at the bottom left and top left of the figure, respectively. The data from the monitoring program MOJAVE at 15 GHz are also available but too far apart in time (4-5 months) to be considered for this rapidly changing source. 

When comparing our observations to the BEAM-ME VLBA data at 43~GHz from close-in-time observations, we can see that not only the general total intensity structure and jet direction remains the same with respect to the one we found, but also the polarization. EVPAs from 43~GHz are found to be oriented along the jet direction on pc scales, indicating a magnetic field mainly perpendicular to the relativistic jet \citep{Jorstad_2005}.

From the EHT image taken 9 months prior to our observations \citep{kim_2020}, we can see how the double structure found in the core fits consistently within the one we found (plotted in contours in the top left panel). The core shift between both frequencies is ignored for the alignment, although it can be estimated as $\sim 40 \, \mu \rm as$. This particular alignment using the brightest pixel is motivated by the incompatible fields of view from EHT (230 GHz) and RadioAstron image (22 GHz). This is due to the different sensitivity to extended structure between the two arrays,
as well as the fact that optically thin jet emission at 22 GHz will most likely be below the flux sensitivity at 230 GHz, even if it was sensitive to these spatial scales. Thus, it makes an alignment on optically thin components unfeasible. 


\begin{figure}
    \centering
    \includegraphics[width=\columnwidth]{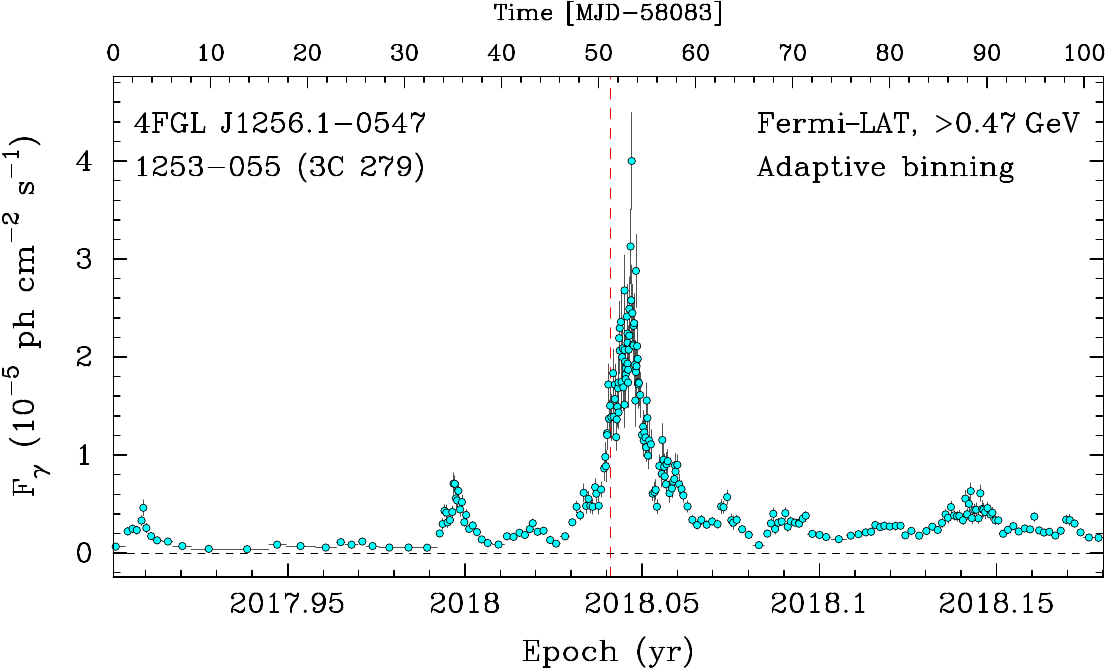}
    \caption{Gamma-ray light curve of 3C\,279 (positionally associated with 4FGL J1256.1-0547) constructed from Fermi-LAT data using adaptive binning \citep{Lott12} with a constant relative uncertainty on flux of $\sim$15\% in each bin. The vertical red dashed line indicates the epoch of \textit{RadioAstron} observations on 15 January 2018, which coincides with the rising phase of a prominent short-duration (6.8 days) flare reaching a peak of $4.0\cdot10^{-5}$ ph~cm$^{-2}$~s$^{-1}$ ($2.4\cdot10^{-8}$ erg~cm$^{-2}$~s$^{-1}$) on January 18, 2018}.
    \label{fig:gamma_lc}
\end{figure}




\begin{figure*}
    \centering
    \includegraphics[trim=0cm 3cm 0cm  2.0cm, clip, width=\textwidth]{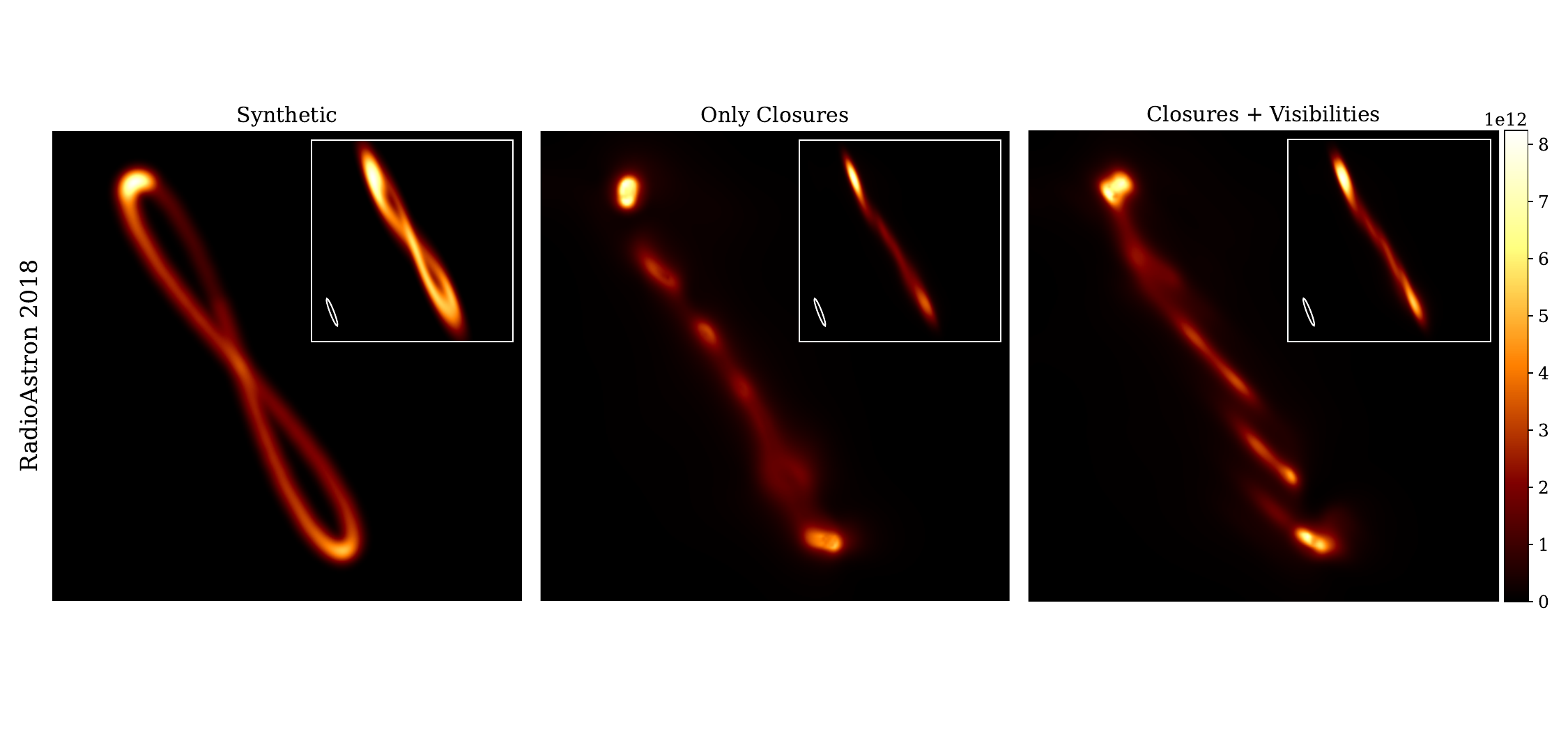}
    \includegraphics[trim=0cm 3cm 0cm  2.0cm, clip,width=\textwidth]{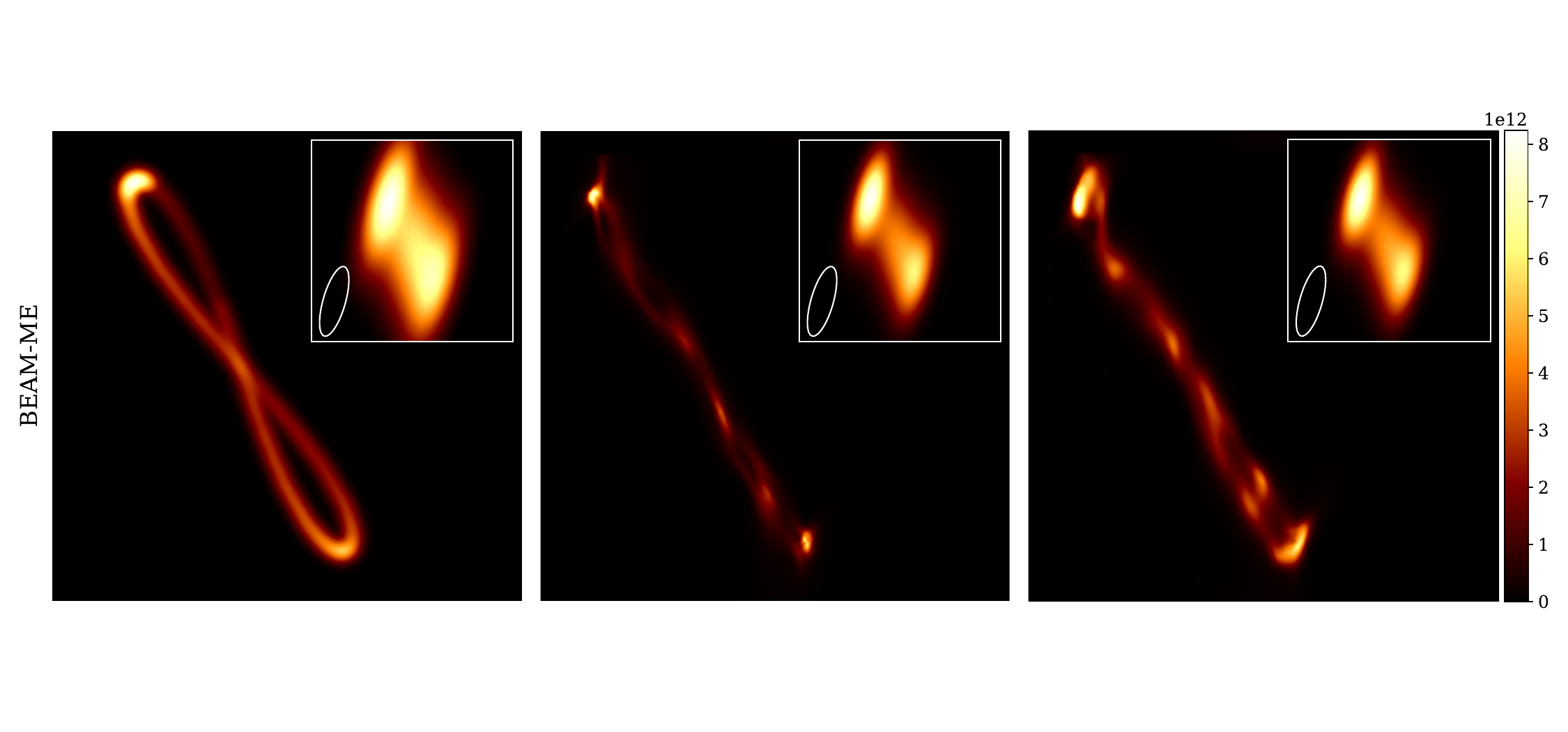}

    \caption{Synthetic data test using different data terms to probe filament detection using the \textit{RadioAstron} $uv$-coverage of 2018 (first row) and BEAM-ME February 2018 (second row). Main images have a field of view of 1.1 $\times$ 1.1 mas, while miniatures at the top right show the main images convolved with its respective nominal beams on a field of view of 1.3 $\times$ 1.3 mas and 2 $\times$ 2 mas, for each row respectively. Units of the color wedges indicate brightness temperature in Kelvin. The nominal beam for each dataset is plotted in white at the bottom left of the miniature images.}
    \label{fig:synthetic}
\end{figure*}

 \subsection{Comparison with the 2014 results} \label{subsec:comparison}

Let us now look at a direct comparison between the previous \textit{RadioAstron} observation of 3C\,279 at 22 GHz from 2014 observations \citep{Fuentes_2023} and our 2018 data. A summary of the main differences is shown in Table \ref{tab:diff}. Fig. \ref{fig:comparison} shows both images overlayed in a $\sim 1.7 \times 1.7 \, \text{mas}$ field of view, where the 2014 epoch is plotted in blue and aligned using the brightest pixel of both images.
The main difference observed in 2018 epoch is that the filamentary structure seen in \citet{Fuentes_2023} (2014 epoch) does not appear to be present four years later, in 2018. The orientation of the jet direction has also changed between both epochs, being slightly rotated in a counterclockwise motion. 

The total intensity of the source has greatly changed, as mentioned in the previous subsection, going from 27.16 Jy in 2014 to 10.2 Jy in 2018. 
In 2014 observations, bright emission regions could be observed due to Doppler boosting. However, these features do not appear so clearly in 2018 data. 
The integrated degree of linear polarization of the source is found to be around 5\% in 2018, unlike the 10\% found in 2014, hinting at more ordered magnetic fields. Nevertheless, from the Fig. \ref{fig:RA_image} it is possible to see that the EVPA direction (see Section \ref{sec:Obs}) is in 2018 similar to 2014 observations, still mostly parallel the jet direction, supporting the existence of a strong toroidal component and in agreement with a helical magnetic field configuration. In some regions, however, such as next to the core or at the bend by the end of the jet we detect, this general direction seems to rotate and become almost perpendicular to it. 
Our findings concur with previous detections of toroidal magnetic fields using VLBI polarimetric studies \citep[e.g.,][]{Cho_2024}, through polarization structures and transverse Faraday rotation measure gradients \citep{toscano_2025, kovalev_2025}.

\subsection{Gamma-ray flaring activity in 2018} \label{subsec:gamma}

As a very active and variable blazar, 3C\,279 is also a powerful gamma-ray source and the first one showing strong and rapid variability at GeV energies \citep{Hartman_1992}. It is also the first flat-spectrum radio quasars (FSRQ) detected above 100 GeV \citep{Magic_2008}.

After several high-energy flares in 2013 and 2015, in 2018, 3C\,279 became very active again, with fluxes exceeding the previous 2015 values. Several studies have been carried out on the January 2018 gamma-ray flare, either focusing on the high-energy correlation between multi-wave bands with hour binned light curves \citep{Shah_2019, Prince_2020, Goyal_2022}, particle acceleration or on magnetic reconnection mechanism \citep{ Wang_2022, Tolamatti_2022}.


The flare reached its peak on Jan 18, 2018 \citep{Shukla2020}, three days after the RA observation. 
The peak of this flare went up to $4.0\cdot10^{-5}$ ph~cm$^{-2}$~s$^{-1}$, which corresponds to 0.0152 $\text{MeV/cm}^2/\text{s}$ ($2.4\cdot10^{-8}$ erg~cm$^{-2}$~s$^{-1}$).
The flare duration inferred as the FWHM of the fitted Gaussian is $\sim$ 6 days.
This event can be compared to two previous major flares of 3C\,279: the June 2015 flare, which peaked at $(3.91\pm0.25)\cdot10^{-5}$ph cm$^{-2}$~s$^{-1}$ \citep{Paliya_2015} with sub-day variability, and the December 2013 flare \citep{Paliya_2016}, characterized by extremely fast and intense sub-flares lasting only a few hours, with a very hard gamma-ray spectrum. While the 2018 flare reached a photon flux comparable to the 2015 event, its smoother and longer evolution stands in contrast to the shorter and more rapidly variable behavior seen in the 2013 and 2015 flares.

\begin{figure}
\centering
    \includegraphics[trim=0 0cm 0 0cm, clip, width=\columnwidth]{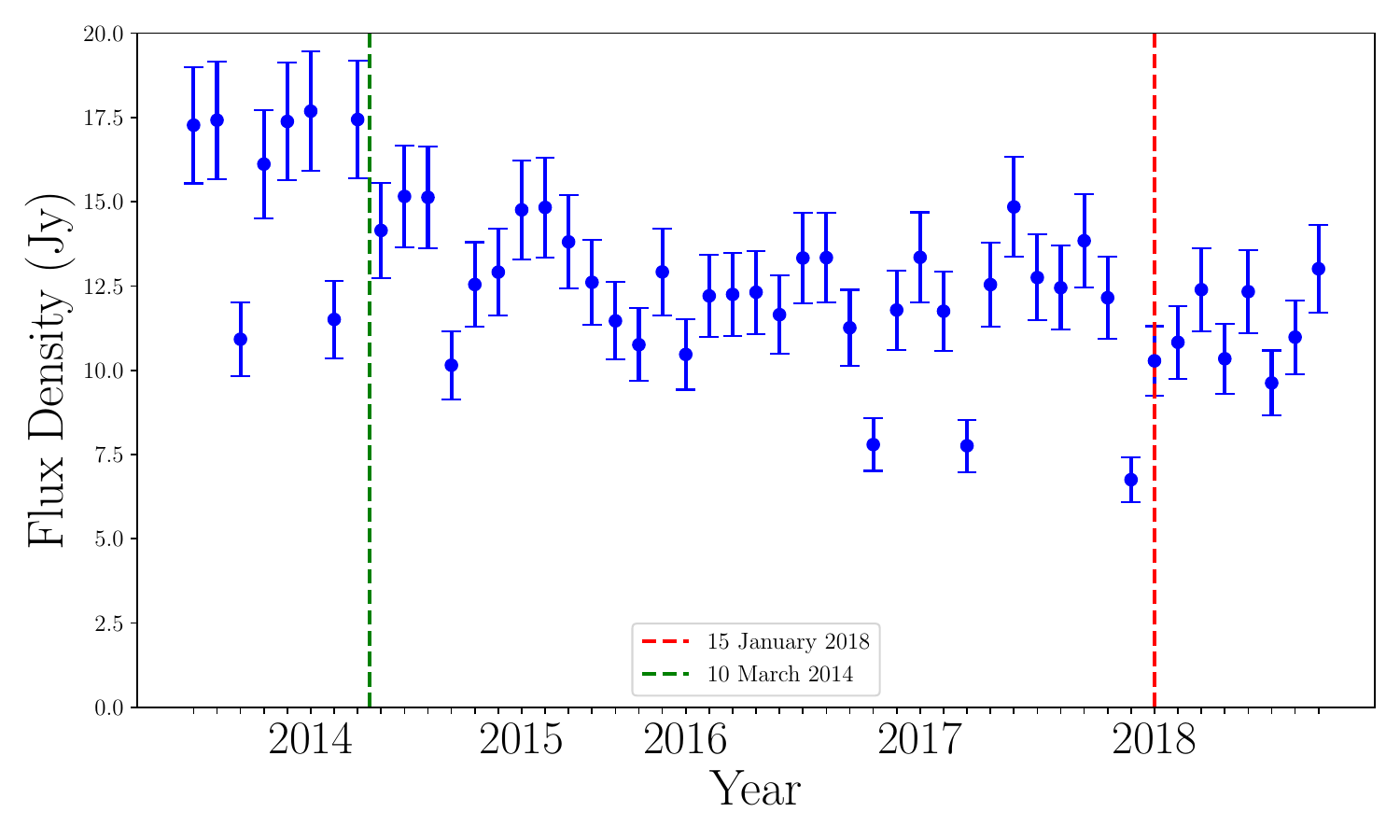}
    \caption{Lightcurve of 3C\,279 at 43~GHz VLBA from the BEAM-ME project constructed using the flux density of CLEAN images (within a field of view of $2 \times 2$ mas). Red line indicated the observations from 2018 and green line from 2014. We estimated data uncertainties of 10\% indicated with error bars.}
    \label{fig:lightcurve}
\end{figure}

Our observations are quite close in time with respect to the gamma-ray flare detected in January~2018. In fact, according to the Fermi database, our observations took place while the flare was in the rising (Fig.~\ref{fig:gamma_lc}). Based on multifrequency analysis presented in \citet{Mohana_2023}, 3C\,279 radio light curve seems quite flat around the moment of our observations (both at 22 GHz and 43 GHz, as can be seen in Fig. \ref{fig:lightcurve}), and does not perceive an increment in flux density until the end of 2020. Thus, the high values reported for the January 2018 gamma-ray flare do not show a visible impact or correlation with our observations, similarly to the flares in 2013 and 2015, which also show no correlation with the radio band.

 \subsection{Synthetic data} \label{subsec:synth}

Previous observations from \citet{Fuentes_2023} were able to detect a filamentary structure. Therefore, it is natural to ask whether the reason for the lack of detection of these filaments in our observations lies behind the natural evolution of the source or elsewhere, such as the sparsity of the uv-coverage.
In order to answer this question, we apply a similar methodology as in \citet{Fuentes_2023} and carry out tests using the same synthetic data model.

The synthetic data in Fig. \ref{fig:synthetic} represent a couple of intertwined filaments forming a rotated-eight shape in a field of view of 1.3 $\times$ 1.3 mas. By using \texttt{eht-imaging} we can simulate observations of the ground truth image under the same conditions of a certain observation. That is to say, using the same baselines, thermal noise, phase corruption, and gain uncertainties from the stations. In this case, we perform this test with data from two different uv-coverages: our epoch from 15 January 2018 at 22GHz and BEAM-ME epoch from 17 February 2018 at 43 GHz, each displayed in a different row. 

For all datasets, the same approach has been used in the imaging reconstruction, using either only closure quantities (closure phases and log closure amplitudes) or a combination of a first image with only closures, then doing self-calibration and adding complex visibilities.

In Fig. \ref{fig:synthetic}, in the first row, we show the results with our \textit{RadioAstron} data. When we use only closure quantities, the reconstruction takes the shape of a roughly straight line in the right direction but shows no sign of filaments. When adding visibilities, the reconstruction marginally improves, but it is still not possible to discern any loops. It is, however, possible to detect the two brightest features at both ends of the loops, though not the one where the filaments intersect. 
Thus, even if filaments were present in the source at the time of our observations, their existence could not be reliably confirmed.

\begin{figure*}[ht]
    \centering
    \includegraphics[trim=0 4cm 0 2cm, clip, width=0.99\textwidth]{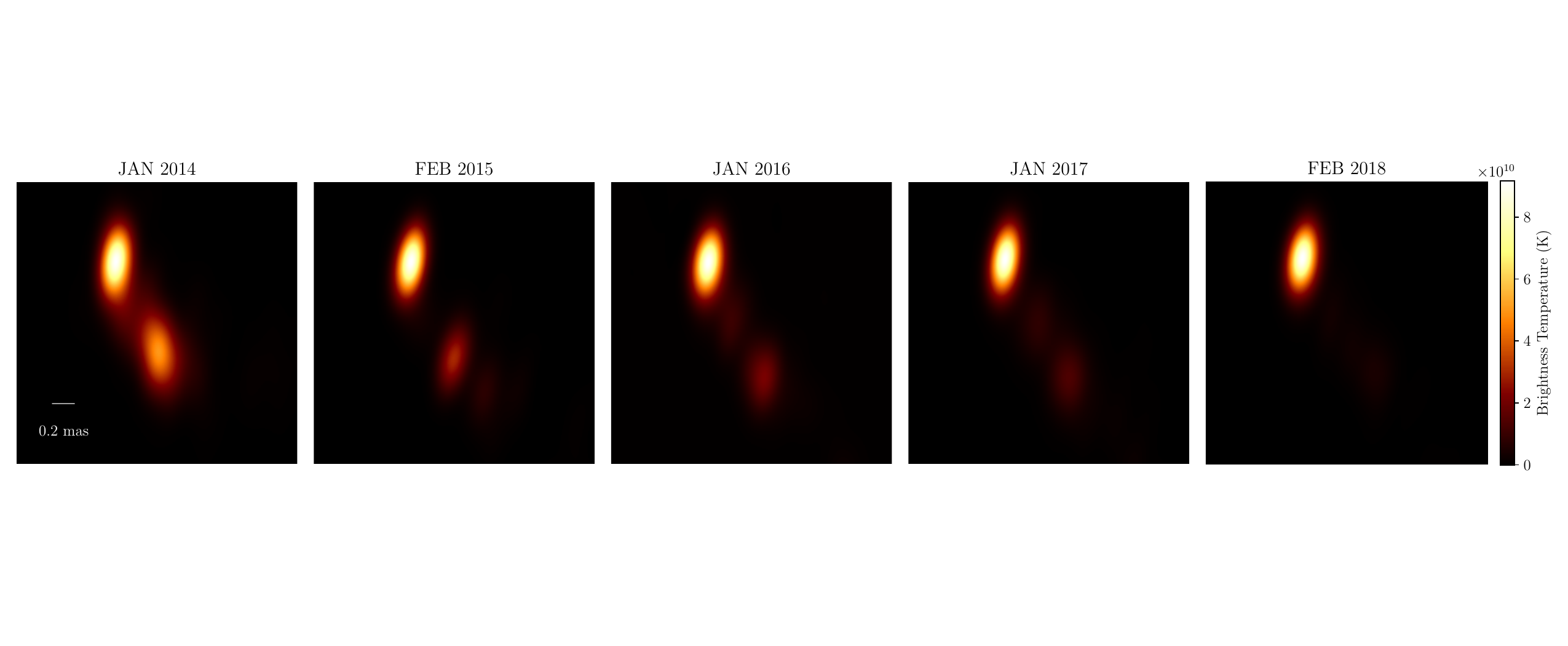}
    \caption{BEAM-ME 43~GHz VLBA yearly evolution from their public data between 2014 and 2018 of 3C\,279 using a field of view of 2 $\times$ 2~mas.}
    \label{fig:beam-me}
\end{figure*}

\begin{figure}
    \centering
    \includegraphics[width=0.7\columnwidth]{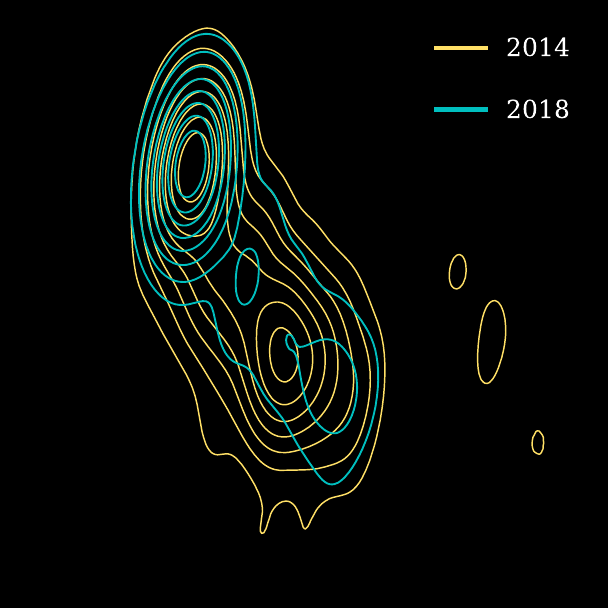}
    \caption{BEAM-ME image comparison between 2014 and 2018 epochs. Contours display 0.2 to 0.9 of the maximum brightness in intervals of 0.1.}
    \label{fig:beammne_22_43}
\end{figure}
For the second row, where the BEAM-ME synthetic test is displayed, we also tried to see if with the coverage of this monitoring program, it would be possible to recover the filamentary structure. Again, as can be observed in Fig. \ref{fig:synthetic} (lower panel), it is not possible to recover any filaments, but an approximately straight line of emission, also showing an uneven brightness distribution. The reconstruction does not really improve when adding the visibilities, as happened in the previous case, but does manage to obtain the bright features at the loop edges.

A successful recovery of the filamentary structure can be seen in the test performed in \citet{Fuentes_2023}, supporting their existence.


\subsection{43~GHz VLBA BEAM-ME evolution} \label{subsec:BEAM}

We can contextualize our results by looking at public data from BEAM-ME at 43~GHz, and thus study the brightness evolution of 3C\,279 jet and its dynamic nature. 
In Fig.~\ref{fig:beam-me}, we show epochs yearly, spanning approximately four years from 2014 to 2018, in order to discern any change in the source that may confirm and help understand the lack of filament detection. 

While in the BEAM-ME webpage there is already access to these images made by Difmap using the traditional CLEAN method, we used the already calibrated fits files and displayed them (not reconstructed) in Stokes I using \texttt{eht-imaging} library, the same one used to reconstruct the 2018 \textit{RadioAstron} data. By doing so, we capture a comprehensive view of the temporal evolution of the blazar over four years, seeing that this period in particular highlights a ``turning off'' phase.

We note that from 2014 to 2018, the source progressively decreased its brightness; this fact 
is also clear if we look at the light curve shown in Fig.~\ref{fig:lightcurve}. This decline in brightness is not uncommon in blazars 
known for the variability of their light curves, often on timescales of years to decades \citep[e.g.,][]{metsahovi_1992,Lister_2018, Mohana_2023, Eppel_2024}. Such variability can be influenced by changes in the Doppler beaming factor or accretion rate onto the black hole. 


Notably, the decrease in brightness of the source poses significant implications for our ability to observe and resolve any structural details in the jet, both in total intensity and polarization. This means that as the source dims towards 2018, features in the jet become harder to detect and any structure is substituted by a mostly straight faint line of emission, impending the study of the intricate magnetic field configurations and plasma dynamics that drive the jet physics in 3C\,279.

Moreover, we can directly compare BEAM-ME images from close-in-time epochs to the \textit{RadioAstron} observations, i.e., from February 2014 and 2018 (see Fig. \ref{fig:beammne_22_43}). It is possible to see 
a slight change in the orientation, with the 2018 epoch showing a small counterclockwise direction. This behaviour is also observed between the \textit{RadioAstron} epochs (see Sec. \ref{subsec:comparison}).




\subsection{Brightness temperature} \label{subsec:Bright_temp}

The plasma acceleration and radiative evolution of the jet's innermost region can be studied through estimations of the brightness temperature \citep{Readhead_1994, Marscher_1995, Kovalev_2005, Fromm_2013, Roder_2025}. It mainly provides an approximation of the source brightness
, making it possible to characterize the energy partition between plasma and the magnetic field assuming self-absorbed synchrotron radiation \citep{Readhead_1994, Homan_2006, Kovalev_2016, Homan_2021}.

\begin{table*}[t]

\centering
\begin{tabular}{cccccc}
\hline\hline
Component & $S$ &                                 FWHM  & r  &  PA &        $T_{\rm b,obs}$ \\

ID & (Jy) & (mas) & (mas) & ($^{\circ}$) & (K) \\
 \hline
Core &            6.40  $\pm$ 0.64  &   0.10 $\pm$ 0.01  &  0                  &    0 
&             (1.6$\pm$ 0.2) $\times 10^{12}$ \\
C1 &              1.04  $\pm$ 0.10  &   0.13 $\pm$ 0.01  &  0.13 $\pm$ 0.01    &    11.3   $\pm$ 1.2  &             (1.5$\pm$ 0.2) $\times 10^{11}$ \\
C2 &              0.29  $\pm$ 0.02  &   0.09 $\pm$ 0.01  &  0.57 $\pm$ 0.03    &    36.9   $\pm$ 3.3  &             (8.8$\pm$ 0.9) $\times 10^{10}$ \\
C3 &              1.58  $\pm$ 0.15  &   0.30 $\pm$ 0.01  &  0.86 $\pm$ 0.02    &    32.3   $\pm$ 3.3  &             (4.3$\pm$ 0.4) $\times 10^{10}$ \\
E  &              0.68  $\pm$ 0.07  &   1.64 $\pm$ 0.02  &  2.26 $\pm$ 0.01    &    36.6   $\pm$ 3.4  &             (6.2$\pm$ 0.6) $\times 10^{8}$\\
\hline
\hspace{1cm}
\end{tabular}
\caption{Model fitting results from the \textit{\textit{RadioAstron}} image. From left to right columns show: Component ID, flux density in Jy, size of the component in mas, radial distance from the core in mas, position angle of the component position with respect to the core in degrees (origin in North and positive counterclockwise direction) and observed brightness temperature without Doppler and redshift correction in Kelvin (K).}
\label{tab:brighttemp}
\end{table*}

The inverse Compton process limits the maximum intrinsic brightness temperature for incoherent synchrotron sources, corresponding to $(0.3$–$1) \times 10^{12}$K \citep{Kellerman_1969}. If a flare has taken place and the brightness temperature exceeds this limit, the source is expected to cool rapidly—on the order of a day—due to catastrophic inverse Compton losses. This estimate assumes a homogeneous, self-absorbed synchrotron source with no significant Doppler boosting, no ongoing particle acceleration, and efficient energy loss via inverse Compton scattering \citep[e.g.,][]{Readhead_1994, Slysh_1992}.
Complimentary to this, another upper threshold arises for sources near the equipartition, referred to as equipartition brightness temperature (T$_\mathrm{eq}$, where the energy density of the magnetic field is approximately equal to the energy density of the relativistic particles), of value $\sim 5\times 10^{10}$ K \citep{Readhead_1994, Singal_2009}.


On another hand, refractive scattering in the interstellar medium can introduce substructure that biases brightness temperature estimates at longer wavelengths \citep{Johnson_2016}, but its impact decreases sharply with frequency. At 22 GHz this effect is negligible and therefore, not taken into account in the following analysis.

The brightness temperature parameter is commonly determined by modeling the structure of the emitting region, estimating the flux density and angular size of the model components.
The observed brightness temperature in this work is calculated by fitting circular Gaussian components with different flux densities and sizes (see Table~\ref{tab:brighttemp} columns 1 to 5 and Fig.~\ref{fig:comparison}) using the \texttt{modelfit} function in Difmap \citep{Pushkarev_2012, Gomez_2022}. The error for the flux density was estimated to be $\sim 10\%$, since it is mostly dominated by the a~priori calibration. Errors for the rest of the quantities were also set to $\sim 10\%$ of their value since the statistical errors from the fitting itself are commonly misleadingly small and do not convey the systematic uncertainties from the measurements. The brightness temperature is given by \citep[e.g.,][]{Gomez_2022}:
\begin{equation}
   \hspace{2cm} T_{\rm b,obs}=1.22 \times 10^{12} \left( \frac{S_{\nu}}{\text{Jy}} \right) \left(\frac{\nu_{\rm obs}}{\text{GHz}}\right)^{-2} \left(\frac{\theta_{\rm obs}}{\text{mas}}\right)^{-2} \, [K]\,,
\label{eq2}
\end{equation}
where $S_{\nu}$ is the component flux density, $\theta_{\rm obs}$ is the size of the model component, and $\nu_{\rm obs}$ is the observing frequency.

Using Eq.~\ref{eq2}, we estimated the brightness temperatures corresponding to the different components of the image model (see Table~\ref{tab:brighttemp}, column~6), finding five different components that amount to 10.0~Jy (out of the 10.2~Jy found in the image), which is a good approximation. 
In order to get a good fit, a large component E is required to capture some extended emission (as seen in Table~\ref{tab:brighttemp}), but it is not shown in Fig.~\ref{fig:comparison}.

Moreover, to calculate the intrinsic brightness temperature of the jet, we also need to take into account the cosmological correction and the Doppler boosting effect. The blazar 3C\,279, with an estimated viewing angle of $\theta \sim 2 ^{\circ}$ \citep{Jorstad_2017}, is known to have a wide range of Doppler factor ($\delta$) values [10-40] \citep{Jorstad_2005,Bloom_2013,Jorstad_2017, kim_2020}. The observed brightness temperature is related to the intrinsic one of the jet $T_{\rm b,int}$ as follows:
\begin{equation}
    \hspace{2.8cm} T_{\rm b,int}= T_{\rm b,obs} \frac{1+z}{\delta}\,,
\end{equation}
where $z$ is the cosmological redshift of the source. 

From the observed brightness temperature values found, assuming the most likely scenario that the radiative losses are from synchrotron radiation, we can see that the observed brightness temperature of the core has a value of $1.6 \times 10^{12}$ K. Such high and even higher values are common in \textit{\textit{RadioAstron}} observations \citep[e.g.,][]{Kovalev_2016,Gomez_2016,Bruni_2017, pilipenko_2017,2020AdSpR..65..705K,Roder_2025}. 

However, the intrinsic brightness temperature value of the core, of $\sim 10^{11} \, \rm K$, calculated assuming a Doppler factor of 24 \citep{Jorstad_2005, kim_2020}, can be reconciled with the Inverse Compton limit, considering the high Doppler factor observed in 3C\,279. This value suggests that the core is close to equipartition, meaning that the energy densities of the relativistic particles and the magnetic field are nearly balanced. This is consistent with the lack of evident flaring events in radio frequencies around the time of our observations. The rest of the components show a decrease of brightness temperature as we move further away from the core, following the expected energy dissipation with distance via adiabatic and radiative losses. 


The observed brightness temperature at the core (including cosmological and Doppler boosting corrections), where the frequency is also affected by Doppler factor so that $ \nu_{\rm int} = \nu_{\rm obs}(1+z)/\delta$, can be related to the magnetic field strength of a synchrotron self-absorbed core \citep[Section 5.3 of][]{Condon_2016}. We can justify the core to be optically thick by establishing a rough spectral index ($\alpha$) comparing the brightness of the 22 GHz \textit{RadioAstron} core (6.40 Jy) to that of the BEAM-ME 43 GHz core (8.85 Jy, calculated also using modelfit), yielding a value of $\sim 0.48$. This result is consistent with optically thick synchrotron emission from a stratified AGN jet core, which typically falls between (0.2-0.6). Therefore, we can apply the magnetic field the relation:
\begin{equation}
   \hspace{1.0cm} B = 1.4 \times 10^{21} \left(\frac{\nu_{\text{obs}}}{\text{GHz}}\right) \left(\frac{T_{\rm b,obs} }{\text{K}}\right)^{-2} \frac{\delta}{(1+z)} \ \text{[G]}\,,
\end{equation}

yielding a magnetic field strength of $\sim $ 0.2~G near the core. This value is consistent with those found in \citet{Roder_2025}, based on Y. Y. Kovalev et al. in prep.

\section{Conclusions}
\label{Sec:Conclusions}

We have carried out \textit{RadioAstron} observations of 3C\,279 on 2018 January 15 at 22~GHz, the last possible of this source with this space VLBI mission. 
The array sensitivity, together with RML methods (\texttt{eht-imaging}), have enabled us to obtain super-resolved images of the 3C\,279 jet with high angular resolution (26 $\mu \rm as$). The images reveal the lack of detection of a filament-like structure found in a similar previous study with RadioAstron, using data from a 2014 epoch \citep{Fuentes_2023}. 

Further analysis using synthetic data with both our array properties and those of BEAM-ME close-in-time, supports the fact that the lack of filament detection is not necessarily related to their non-existence, but most likely due to an insufficient uv coverage.
Moreover, observations from BEAM-ME public data reveal that over the 4-year-period 2014-2018, the source exhibits a gradual decline in brightness, affecting our ability to resolve its jet structure in both total intensity and polarization.

We also explore the gamma-ray emission relation with our observations, finding that they coincide with the rising of a gamma-ray flare surpassing the last one in brightness from 2015. Nevertheless, we find that this phenomenon does not seem to impact or correlate with our results.

Finally, a study of the brightness temperature using model fitting with Gaussian components yields a core with an observed brightness temperature of the order of $10^{12}$\,K, in agreement with previous \textit{RadioAstron} observations, and an estimated magnetic field strength of $\sim 0.2$ G in the core region. The values found for the jet's brightness temperature suggest it to be close to equipartition, as can be expected for a source in a low activity period.



\begin{acknowledgements}

The work at the IAA-CSIC is supported in part by the Spanish Ministerio de Econom\'{\i}a y Competitividad (PID2022-140888NB-C21), the Ramón y Cajal grant RYC2023-042988-I from the Spanish Ministry of Science and Innovation, the Consejo Superior de Investigaciones Cient\'{\i}ficas (grant 2019AEP112), and the State Agency for Research of the Spanish MCIU through the ``Center of Excellence Severo Ochoa" grant CEX2021-001131-S funded by MCIN/AEI/ 10.13039/501100011033 awarded to the Instituto de Astrof\'{\i}sica de Andaluc\'{\i}a.

LIG gratefully acknowledges support by the Chinese Academy of Sciences PIFI program, grant No.~2024PVA0008. 
%

YYK was supported by the MuSES project, which has received funding from the European Union (ERC grant agreement No 101142396). Views and opinions expressed are however those of the author(s) only and do not necessarily reflect those of the European Union or ERCEA. Neither the European Union nor the granting authority can be held responsible for them. 


\textit{RadioAstron}, a project led by the Astro Space Center of the Lebedev Physical Institute of the Russian Academy of Sciences and the Lavochkin Scientific and Production Association under a contract with the Russian Federal Space Agency, in collaboration with partner organizations in Russia and other countries. 

The European VLBI Network, a joint facility of independent European, African, Asian, and North American radio astronomy institutes. Scientific results from data presented in this publication are derived from the EVN project. 

This research is partly based on observations with the 100m telescope of the MPIfR at Effelsberg. 

This publication makes use of data obtained at Metsähovi Radio Observatory, operated by Aalto University in Finland. 

The Long Baseline Array is part of the Australia Telescope National Facility (https://ror.org/05qajvd42) which is funded by the Australian Government for operation as a National Facility managed by CSIRO.

Special thanks go to the people designing and supporting the space and ground observations. 

This research is based on observations correlated at the Bonn Correlator modified for \textit{RadioAstron} correlation, jointly operated by the Max-Planck-Institut für Radioastronomie, and the Federal Agency for Cartography and Geodesy. 

This study makes use of 43 \, GHz VLBA data from the VLBA-BU Blazar Monitoring Program (VLBA-BU-BLAZAR; \url{http://www.bu.edu/blazars/BEAM-ME.html}), funded by NASA through Fermi Guest Investigator grant number 80NSSC20K1567.

\end{acknowledgements}

\bibliography{references}

\bibliographystyle{aa}

\appendix
\onecolumn

\section{Difmap model image}
\label{appendixa}
\begin{figure}[ht]
    \centering
    \includegraphics[width=0.4\textwidth]{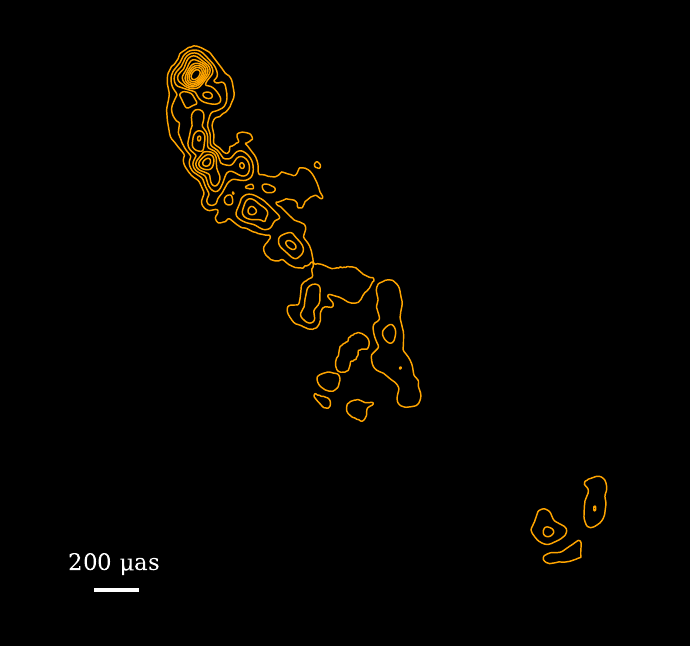}
    \caption{CLEAN model in contours of 3C\,279 \textit{RadioAstron} 2018 observations using Difmap and plotted using a Python script. Contours are plotted at levels of 0.1 to 0.9 times the peak brightness, in steps of 0.1.}
    \label{fig:difmapmod}
\end{figure}

\section{Parameter survey}\label{appendixb}

To ensure a robust reconstruction regardless of the regularizers used in the imaging process, we perform an additional parameter survey. The values selected survey over lower values of 'mem' parameter, testing the impact of the prior image resemblance, and higher values for 'tv' and 'l1', checking that neither the regularizers enhancing smoothness or compactness, respectively, cause any major structural changes in the final reconstruction.

\begin{figure}[ht]
    \centering
    \includegraphics[trim=2cm 6cm 2cm 2cm, clip, width=0.82\textwidth]{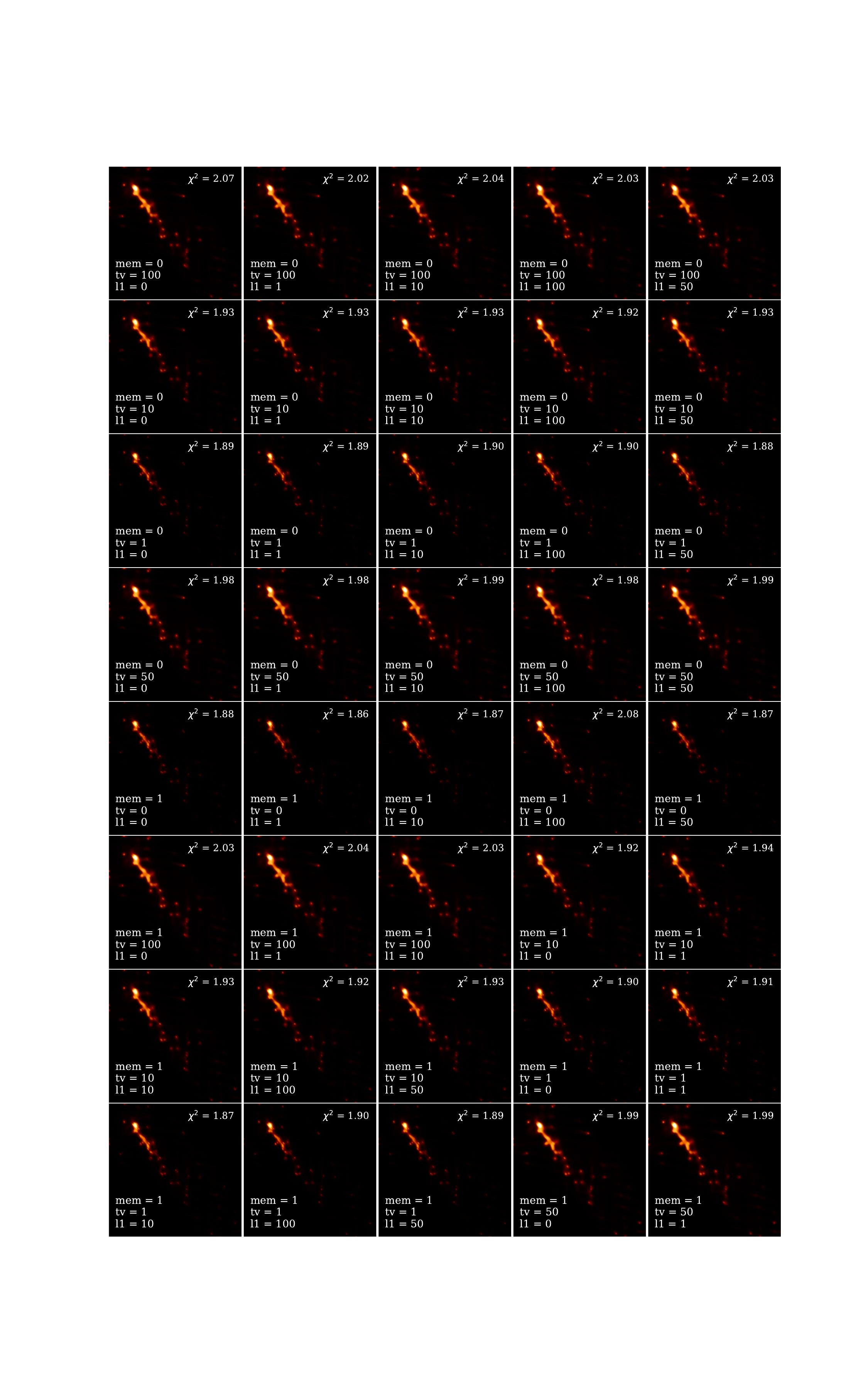}
    \caption{Parameter survey of 3C\,279 \textit{RadioAstron} 2018 displaying similar morphology. Visibilities $\chi^2$ values are at the top right of each image, and the regularizer combination at the bottom left.}
    \label{fig:paramsurvey}
\end{figure}
\end{document}